\documentclass[grl]{agutex2arxiv}
\usepackage{amsmath,amssymb}
\usepackage{graphicx,color}
\usepackage{lineno}
\usepackage{lipsum}
\setkeys{Gin}{draft=false}

\authorrunninghead{PASSEL\`EGUE ET AL.}
\titlerunninghead{FAULT REACTIVATION}

\authoraddr{F. X. Passel\`egue, passelegue@gmail.com}

\begin{document}

\title{Fault reactivation by fluid injection: Controls from stress state and injection rate}

 \authors{Fran\c cois. X. Passel\`egue,\altaffilmark{1} Nicolas Brantut,\altaffilmark{2} Thomas M. Mitchell,\altaffilmark{2}}

\altaffiltext{1}{Laboratoire Exp\'erimental de M\'ecanique des Roches, \'Ecole Polytechnique F\'ed\'erale de Lausanne, Lausanne, Switzerland.}

\altaffiltext{2}{Department of Earth Sciences, University College London, London, United Kingdom.}

\begin{abstract}
  We studied the influence of stress state and fluid injection rate on the reactivation of faults. We conducted experiments on a saw-cut Westerly granite sample under triaxial stress conditions. Fault reactivation was triggered by injecting fluids through a borehole directly connected to the fault. Our results show that the peak fluid pressure at the borehole leading to reactivation depends on injection rate. The higher the injection rate, the higher the peak fluid pressure allowing fault reactivation. Elastic wave velocity measurements along fault strike highlight that high injection rates induce significant fluid pressure heterogeneities, which explains that the onset of fault reactivation is not determined by a conventional Coulomb law and effective stress principle, but rather by a nonlocal rupture initiation criterion. Our results demonstrate that increasing the injection rate enhances the transition from drained to undrained conditions, where local but intense fluid pressures perturbations can reactivate large faults.

\end{abstract}

\begin{article}
  
\section{Introduction}

In the last decade, exploitation of geothermal and hydrocarbon reservoirs \citep{Warpinski1987,cornet1997}, as well as deep fluid injection for geological storage \citep{healy1968,raleigh1976,Zoback1997} induced a strong increase in seismicity \citep{Ellsworth2013}. One of the strongest evidence for injection-induced seismicity is the recent rise in the earthquake rate in Oklahoma that occurred since the beginning of deep waste-water injection associated with unconventional oil reservoir exploitation \citep{Ellsworth2013}. This induced seismicity includes earthquakes of magnitude larger than 5 that have caused significant damage \citep{Keranen2013,Rubinstein2014}. The recent Pawnee $M_w$=5.8 earthquake was the largest in instrumented history in Oklahoma and is only slightly below the maximum magnitude expected during 1900 years of tectonic activity \citep{Langenbruch2016}. 
In order to reduce seismic hazard, Oklahoma regulators have planned a 40 percent reduction in the injection volume per day. However, the influence of injection rate on fault reactivation and induced seismicity remains poorly documented and the result of this decision remains uncertain \citep{Dieterich2015,Langenbruch2016,Goebel2017,Barbour2017}.


From a physical point of view, our understanding of the mechanics of fault reactivation and earthquake nucleation due to fluid pressure variations is based on the concept of effective stress combined with a Coulomb failure criterion. The onset of fault reactivation is typically characterised by a critical shear stress $\tau_\mathrm{p}$ given by the product of a friction coefficient $\mu$ (ranging from 0.6 to 0.85 in most crustal rock types, see \citet{Byerlee1978}) and the normal stress $\sigma$ applied on the fault. In the presence of fluids, this normal stress is offset by an amount equal to the fluid pressure $p$, so that the fault reactivation criterion is \citep{Sibson1985a,Jaeger2009}
\begin{linenomath}
  \begin{equation}\label{eq:tau}
    \tau\geq\tau_\mathrm{p}=\mu(\sigma-p).
  \end{equation}
\end{linenomath}
This simple concept has been used extensively to explain a range of natural and experimental rock deformation phenomena \citep[see reviews in ][]{Scholz2002,Paterson2005}.

However, the reactivation criterion based on the effective stress law is expected to hold (within a reasonable degree of approximation) only when the entire fault is affected by fluid pressure, i.e., if $p$ is homogeneous. In other words, the criterion \ref{eq:tau} is best viewed as a local one, and the onset of large scale fault motion depends on the distribution of fluid pressure, applied stresses and elastic stress redistribution due to partial slip.

The reactivation of slip and the mode of sliding (either quasi-static or dynamic) produced by fluid pressure perturbations has been studied extensively in theoretical models based on fracture mechanics \citep{Viesca2012,Garagash2012,Galis2017}. These approaches show that local fluid overpressures (i.e., $p$ locally greater than expected from a homogeneous Coulomb criterion) can lead to periods of quasi-static, partial fault slip, followed by earthquake nucleation and propagation well beyond the initial pressurised area. Such predictions are in qualitative agreement with field observations showing induced seismicity and fault reactivation in crystalline basements, far from the injection sites \citep[e.g.][]{Keranen2013}. However, theoretical models are necessarily based on simplified assumptions regarding friction laws and do not systematically account for potential couplings between fluid pressure, hydraulic and mechanical properties of rocks and fault interfaces. The reactivation of faults by fluid pressure variations is expected to be complicated by these coupled hydromechanical processes, and accurate predictions at field scale require an in-depth knowledge of the key controlling fault zone properties and injection parameters.

Here, we investigate the conditions for and the characteristics of fault reactivation due to fluid injection in controlled laboratory experiments. We specifically test how the injection rate and background stress conditions influence the onset of fault reactivation. We performed injection tests on saw-cut Westerly granite samples subjected to triaxial stress conditions, and monitored contemporaneously the evolution of fault slip, stress and elastic wave velocities across the fault. This setup allowed us to analyse \emph{in situ} the effect of fluid pressure diffusion on fault reactivation under realistic upper crustal conditions in a major crystalline basement rock type.

\section{Experimental setup and methods}

\begin{figure*}
\begin{center}
\includegraphics{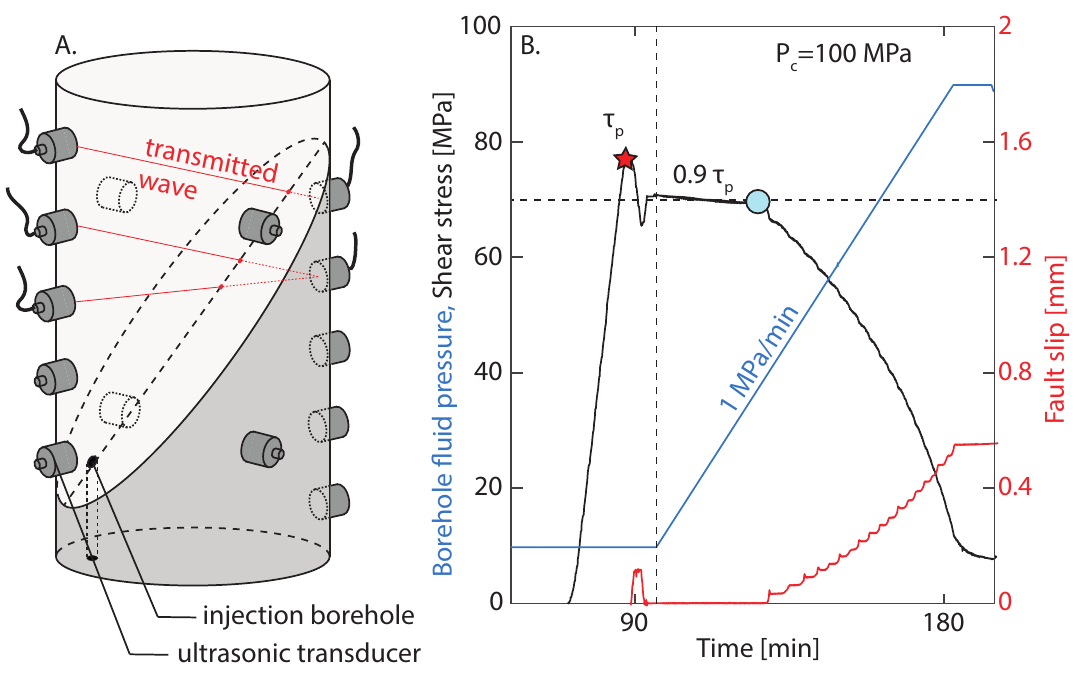}
\end{center}
\caption{Experimental setup. (A) Schematic of the sample assembly. The length of the fault is 8 cm along strike. Injection is conducted in the bottom sample through a borehole reaching the fault surface. (B) Fluid pressure, shear stress and slip measured during a sliding test at constant pressure (left of dotted vertical line) and during a fluid injection test at 1 MPa/min and initial shear stress equal to 90\% of the static frictional strength.}
\label{fig:setup}
\end{figure*}

A cylindrical sample of Westerly granite of 40~mm in diameter was cored, and then cut and precisely ground to a length of 100~mm. The cylinder was then cut at an angle of 30$^\circ$ with respect to its axis of revolution to create an elliptical saw cut fault interface (Figure \ref{fig:setup}A) of 40~mm in width and 80~mm in length along strike. The fault surface was prepared with a surface grinder. A 4~mm diameter borehole, the centre of which was located at 4.5~mm from the edge of the cylinder, was drilled through the material on one side of the fault, connecting the fault surface to the bottom end of the sample (Figure \ref{fig:setup}A).

The faulted sample was placed in a viton jacket, and equipped with 14 piezoeletric transducers arranged in an array shown schematically in Figure \ref{fig:setup}A. Each transducer consists in an aluminium casing that embeds a 1~mm thick, 5~mm diameter piezoelectric ceramic disk (material reference PIc255 from Physik Instrumente GmbH) that is polarised perpendicular to the sample surface. Five pairs of transducers were positioned along the cylinder in the plane perpendicular to fault strike. This two dimensional array allowed us to monitor wave velocity variations along 25 ray paths intersecting the fault at a range of locations. Two additional pairs of transducers were placed at the top and bottom of the sample at 90$^\circ$ to the main array.

The instrumented sample was placed in the 400 MPa triaxial oil-medium apparatus of the Rock and Ice Physics Laboratory at University College of London \citep{Eccles2005}. The bottom end of the sample, where the borehole is located, was connected to a high-capacity servo-hydraulic pore fluid intensifier instrumented with a pressure transducer and an LVDT (Linear Variable Differential Transducer) that measures the variations of the intensifier fluid volume. The top part of the sample was connected to a closed reservoir instrumented with a separate pressure transducer. The pore fluid used in this study was distilled water. The confining pressure ($P_\mathrm{c}$) and the axial differential stress ($Q$) were controlled independently by an electromechanical pump and a servo-hydraulic actuator, respectively. Sample shortening was calculated from an external measurement of the ram displacement, corrected from the stiffness of the loading column. Axial load was measured using an external load cell, and corrected for seal friction. The differential stress on the sample is computed as the ratio of corrected load over sample cross-sectional area. Fault slip is computed by projecting the sample axial shortening onto the fault direction. The average fault normal and shear stresses are obtained by resolving the triaxial stress state onto the fault plane.

During experiments, ultrasonic wave velocities were measured repeatedly in the following manner. An elastic wave was generated at a known origin time by imposing a high voltage ($\sim$250~V), high frequency (1~MHz) electric pulse on a given piezoelectric transducer, and the resulting signals were amplified at recorded (at a 50~MHz sampling frequency) on the 13 remaining sensors. This procedure was repeated sequentially so that all transducers are used as active sources, thus generating a total 14$\times$13 waveforms (hereafter called a ``survey''). During postprocessing, a reference survey is chosen and arrival times of ballistic P-waves are picked manually on all available waveforms. A cross-correlation procedure is employed to determine accurate relative variations in arrival times relative to the reference survey \citep[see details in][]{Brantut2015}. The relative change in wave velocity between each pair of sensors is obtained as the ratio of the change in arrival time over the reference arrival time, and we also correct from the change in relative position of the sensors as the fault slides.

We conducted experiments at two confining pressures, 50 and 100~MPa. The initial pore pressure was set to 10~MPa. The shear stress at the onset of fault slip under constant pore pressure conditions, denoted $\tau_\mathrm{p}$, was determined by conducting an axial loading test. Subsequently, the load was decreased down to a given initial stress $\tau_0$, and the actuator position was maintained constant by a servo-controlled loop on the external displacement transducers. This situation corresponds to a ``stress relaxation'' test, whereby a finite amount of elastic strain energy is stored in the loading column, and any shortening of the sample (here, slip on the fault) is accompanied by a decrease in the applied stress, in a constant proportion of the sample shortening determined by the machine stiffness. This method ensured that fault slip could not runaway beyond a manageable quantity, while not precluding in principle the occurrence of stick-slip events. Fluid was then injected through the borehole at a constant pressure rate (from 1 to 1000~MPa/min, measured at the outlet of the pore fluid intensifier), up to a target value of 40~MPa and 90~MPa at $P_\mathrm{c}=50$~MPa and $P_\mathrm{c}=100$~MPa, respectively. The permeability of the westerly granite is increasing from $10^{-22}$ to $10^{-20}$ in the range of effective confining pressure tested (from 90 to 10 MPa effective confining pressure)\citep{Nasseri2009,Rutter2017}. During injection, ultrasonic surveys were performed at $\sim$10~s time intervals, and other mechanical data were recorded at $\sim$5~Hz.

\section{Mechanical results}

A representative example of shear stress, borehole fluid pressure and slip evolution is shown in Figure \ref{fig:setup}B, for a test conducted at $P_\mathrm{c}=100$~MPa. During the first stage, the onset of fault slip was measured at $\tau_\mathrm{p}=78$~MPa. The stress was then decreased to $\approx0.9\times \tau_\mathrm{p}$, and fluid injection was conducted at a rate of 1~MPa/min. The initiation of fault slip was detected at a borehole fluid pressure of $P_\mathrm{f,bore} \approx 40$~MPa. With further fluid pressure increase, fault slip continues in a series of steps of $\sim$20~$\mu$m in amplitude and $\sim$30~s in duration, separated by dwell times of the order of 200~s. A similar behaviour is observed in all tests.

\begin{figure*}
\begin{center}
  \includegraphics{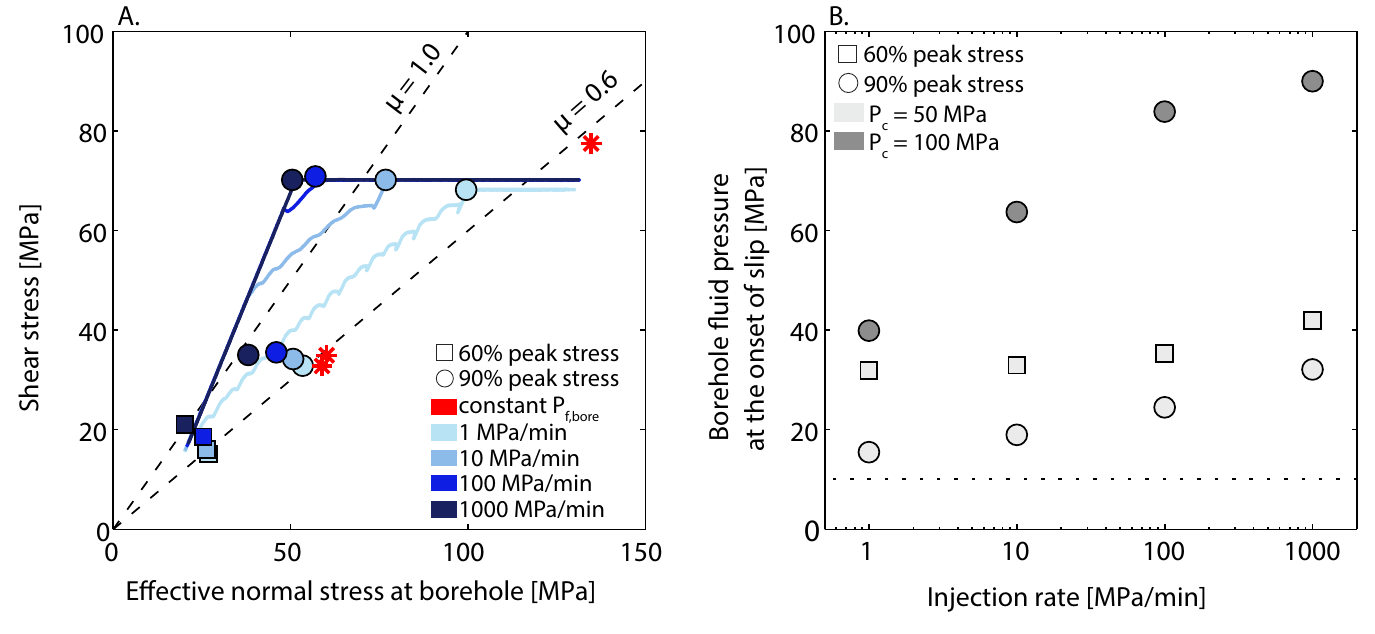}
\end{center}
\caption{Static reactivation of the experimental fault. (A) Mohr diagram presenting the fault reactivation conditions in all experiments. Solid lines correspond to the tracks of the stresses obtained during the experiment conducted at 100 MPa confining pressure and initial stress equal to 90\% of $\tau_\mathrm{p}$. Dashed lines correspond to frictional strength using friction coefficients of 0.6 and 1. Symbols correspond to the state of stress at the onset of slip for each experiments and conditions tested. Red stars: onset of slip under constant pressure conditions. Squares: onset of slip during injection at an initial stress equal to 60\% of $\tau_\mathrm{p}$. Circles: onset of slip during injection at an initial stress equal to 90\% of $\tau_\mathrm{p}$. (B) Fluid pressure allowing fault reactivation as a function of the injection rate. The dashed line corresponds to the background level of the fluid pressure in our experiments.}
\label{fig:reactivation}
\end{figure*}

The complete stress paths of injection tests performed at $P_\mathrm{c}=100$~MPa are shown in Figure \ref{fig:reactivation}A in effective normal stress (considering the \emph{borehole} fluid pressure $P_\mathrm{f,bore}$) vs. shear stress space. The onset of fault slip is also marked for all experiments. Fluid injection leads to a decrease of effective normal stress at the borehole without change in the background shear stress. Once the state of stress reaches a critical point (cf. squares and circles), slip initiates and the values of both shear stress and effective normal stress decline progressively. In the tests conducted under the lowest stress conditions ($P_\mathrm{c}=50$~MPa and $\tau_0=0.6\tau_\mathrm{p}$) and for injection rates up to 100~MPa/min, the onset of fault slip occurs at an effective normal stress comparable to that expected from a static friction criterion with $\mu\approx0.7$, compatible with the static friction of $\mu=0.6$ observed in the test at constant fluid pressure. At 1000~MPa/min injection rate, the fault reactivates at a slightly lower effective normal stress (as measured at the borehole). When the initial stress $\tau_0$ is around 90\% of $\tau_\mathrm{p}$, the effective normal stress required to activate fault slip is generally higher than for low initial stresses. At 1~MPa/min, the fault is reactivated at $P_\mathrm{f,bore}=15.5$~MPa, consistent with the measured static friction. However, with increasing injection rate, the effective normal stress at reactivation decreases significantly. The same trend is observed in the tests performed at $P_\mathrm{c}=100$~MPa.

The borehole fluid pressure required to reactivate the fault is plotted as a function of injection rate in Figure \ref{fig:reactivation}B. For all the tests conducted at elevated initial stress, a clear trend towards high fluid pressure at reactivation is observed as the injection rate increases. This trend is clearest at $P_\mathrm{c}=100$~MPa, where an increase in injection rate from 10 to 100~MPa/min produces an increase in fluid pressure at reactivaton from around 64 to 84~MPa. At the highest injection rate (1000~MPa/min), the target pressure of 90~MPa was reached and slip initiated 4.8~s after that point. The stress paths (as measured at the borehole) shown in Figure \ref{fig:reactivation}A are significantly above the static friction criterion, especially at high injection rate.

\section{Wave velocity variations}

Our experimental results suggest that at low injection rate and low stress, the onset of fault reactivation follows a regular Coulomb criterion. However, increasing the injection rate trends to modify the reactivation criterion of the same experimental fault. This behaviour is enhanced by increasing the state of stress acting on the fault plane prior the injection.

\begin{figure*}
\begin{center}
\includegraphics{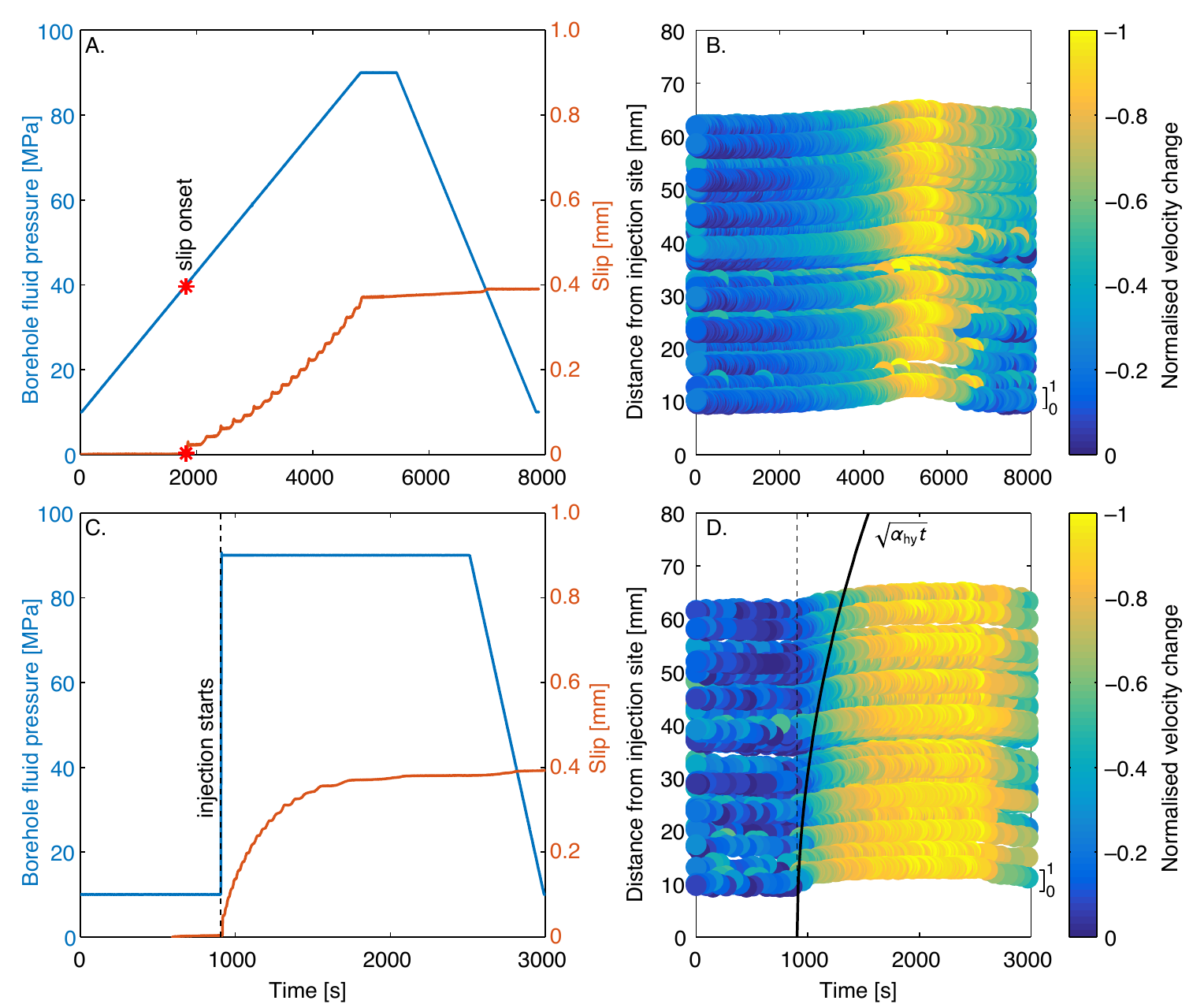}
\end{center}
\caption{Fluid pressure and wave velocities during injection. (A). Evolution of fluid pressure and slip during injection conducted at 1~MPa/min, $P_\mathrm{c}=100$~MPa. (B) Time evolution of the normalised change in compressional wave velocity during injection conducted at 1~MPa/min, $P_\mathrm{c}=100$~MPa, as a function of distance between injection point and intersection point of raypath and fault plane. The velocity along each path is normalised by the difference between its extrema. Both vertical variation and colorbar corresponds to the normalised velocity change. (C) and (D) are similar to (A) and (B) for an injection conducted at 1000~MPa/min.}
\label{fig:wavevel}
\end{figure*}

To understand this change in the onset of fault reactivation with increasing injection rate, we use elastic wavespeed measurements as a proxy to track the evolution of the fluid pressure along the fault. Increasing the fluid pressure under constant confining pressure conditions is expected to decrease of the elastic wavespeed of the fault system as a response to (i) microcracks opening due to the decrease of the effective pressure in the bulk material \citep{Walsh1965,Nasseri2009} and (ii) the decrease of the contact stiffness of the experimental fault \citep{Gueguen2011,Kelly2017}. The evolution of compressional wave velocities across the fault during injections conducted at 1 and 1000~MPa/min and 100~MPa confining pressure are presented in Figure \ref{fig:wavevel}.

During the injection conducted at 1~MPa/min (Figure \ref{fig:wavevel}A,B), the wave velocity along each raypath decreases progressively with increasing fluid pressure, down to a maximum drop of the order of 2--3\% at the maximum fluid pressure of 90~MPa (10~MPa effective confining pressure). At this injection rate, the velocity evolves homogeneously independently of the distance from the injection site, which suggests that fluid pressure is relatively homogeneous along the fault throughout the injection process. 

At high injection rate (1000~MPa/min, Figure \ref{fig:wavevel}C,D), the drop in velocity is a function of the distance from the injection site. At early stage, when the fluid pressure has reached its maximum of 90~MPa and fault slip initiates, a sharp drop in velocity is observed along the raypaths crossing the fault nearest to the injection site, while no change is observed along raypaths intersecting the fault at distances larger than 3~cm from the injection site. Far from the injection point, the velocity change is gradual, and reaches its maximum $\sim$500~s after injection starts. Considering that the changes in wave velocities observed along the fault are mostly due to the propagation of a fluid pressure front during fluid injection, we compute an estimate of the location of the diffusive pressure front as $\sqrt{\alpha_\mathrm{hy}t}$, where $\alpha_\mathrm{hy}$ is the hydraulic diffusivity of the fault and $t$ is the time from the beginning of injection. Using $\alpha_\mathrm{hy}=10^{-5}$~m$^2$s$^{-1}$, the time at which 80\% of the maximum drop in wave velocity is observed along each raypath is acceptably matched by the characteristic dffusion time from the pressurised borehole (see black line in Figure \ref{fig:wavevel}D).

\section{Influence of background stress level and injection-rate on fluid pressure heterogeneity}

Our experimental results indicate that while fluid pressure remains homogeneous over the fault during experiments conducted at low injection rate, a fluid pressure front is observed at high injection rate. This suggests that increasing the fluid injection rate leads to an increase of the fluid pressure heterogeneity over the fault plane. Note that the background stress prior injection, as the confining pressure acting on the fault system, seems to enhance this heterogeneity, i.e., the intensity of fluid overpressure leading to instability (Figures \ref{fig:reactivation}A). To analyse these processes further, we follow \citet{Rutter2017} and compute a pore pressure excess factor defined as
\begin{equation}
  \label{eq:px}
  p_\mathrm{x}=\frac{P_\mathrm{f,bore}}{\sigma_\mathrm{n}-\tau/\mu}
\end{equation}
where $\mu=0.6$ is the static friction coefficient of the fault. A local fluid overpressure is required to reactivate the fault if $p_\mathrm{x}$ becomes greater than 1 during the fluid injection history, whereas $p_\mathrm{x}=1$ if the effective stress principle applies. The evolution of $p_\mathrm{x}$ as a function of the shear stress during each injection is presented in Figure \ref{fig:overpressure}. During the tests conducted at $P_\mathrm{c}=50$~MPa and $\tau_0=0.6\tau_\mathrm{p}$, the fault generally reactivates following $p_\mathrm{x}\approx 1$. Only the injection conducted at the highest rate (1000~MPa/min) highlights fluid overpressure, with a pore fluid excess factor of around 1.4 at the onset of fault reactivation (Figure \ref{fig:overpressure}). By contrast, at elevated background stress and confining pressure ($P_\mathrm{c}=100$~MPa, $\tau_0=0.9\tau_\mathrm{p}$), $p_\mathrm{x}$ increases significantly with increasing injection rate, from 1.5 to 3.7 as injection rate increases from 1 to 1000~MPa/min. Overall, increases in either confining pressure or initial shear stress lead to an increase of the local fluid pressure required for fault reactivation. Beyond the onset of slip, the fault offloads due to the progressive relaxation of the loading column as slip proceeds, and the amount of required fluid overpressure decreases. This observation is similar to that of \citet{Rutter2017} during fluid pressurisation of faults with low hydraulic conductivity.

\section{Discussion and Conclusions}

Our results show unambiguously that both stress state and injection rate modify the conditions for fault reactivation. At elevated stress and during fast, \emph{local} fluid injection, our wave velocity measurements indicate that fluid pressure is heterogeneous along the fault, which explains why a conventional Coulomb law combined with a simple effective stress law is not an appropriate reactivation criterion. In other words, we observe a transition from a homogeneous case, where a static empirical friction law applies (within typical uncertainties), to a heterogeneous case, where the onset of sliding is best viewed as a (nonlocal) fracture problem and controlled by local stress concentrations along the fault plane \citep[e.g.,][]{Rubinstein2004,Svetlizky2014}.

The degree of heterogeneity in pore fluid pressure along the fault is controlled by the balance between the hydraulic diffusion rate and injection rate. The hydraulic properties of the fault and surrounding rock are thus key parameters controlling the extent to which fault reactivation deviates from Coulomb's law. At a given injection rate, our results show that an increase of the confining pressure and/or initial shear stress enhances the fluid pressure required at the injection site for fault reactivation, as well as the fluid pressure heterogeneity along the fault. This observation is explained by the fact that increasing fault normal and shear stresses tend to reduce fault hydraulic transmissivity\citep[e.g.,][]{Rutter2017}, and overall increases in mean stress reduce bulk rock permeability \citep[e.g.,][]{Nasseri2009}. This decrease in hydraulic transmissivity slows fluid diffusion along the fault, and promotes the transition from ``drained'' to ``undrained'' conditions. In the former case, fluid pressure remains homogeneous, whereas in the latter case a fluid pressure front is formed within the finite length of our experimental fault.

\begin{figure*}
\begin{center}
\includegraphics{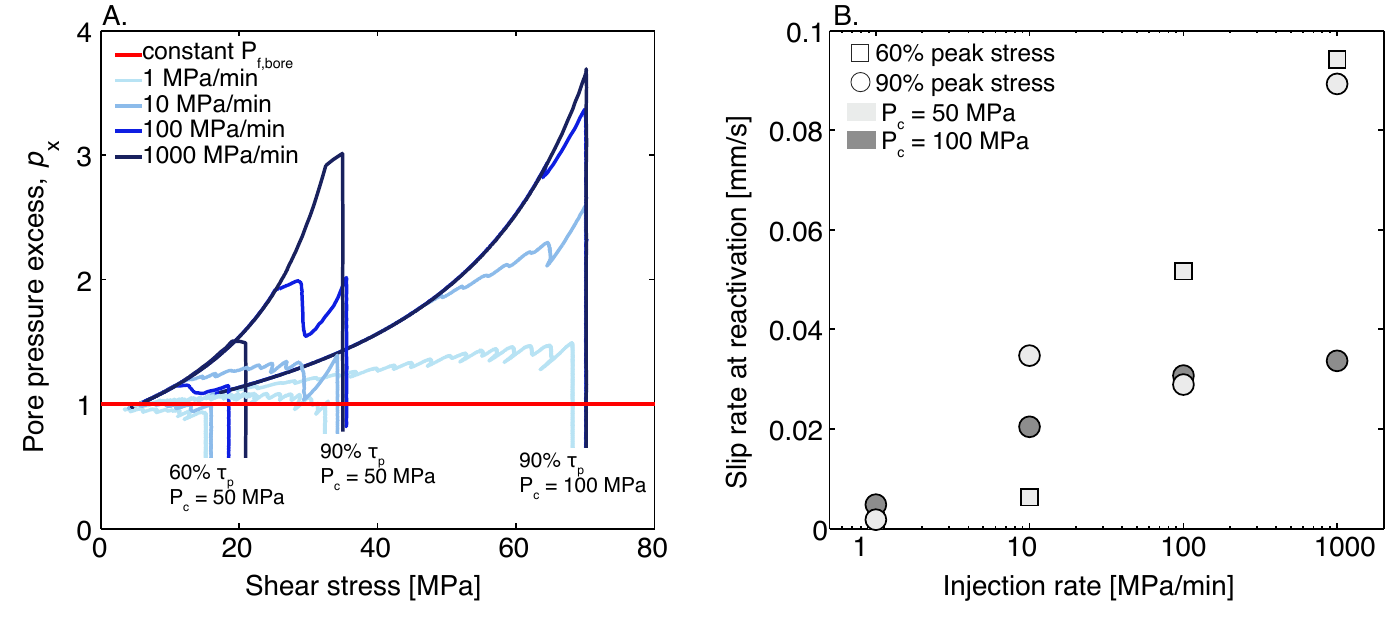}
\end{center}
\caption{Fluid overpressure and slip rate during reactivation. (A) Evolution of the pore pressure excess factor (Equation \protect\ref{eq:px}) during each fluid injection. (B) Slip rate at the onset of slip as a function of the injection rate.}
\label{fig:overpressure}
\end{figure*}

In our experiments, the onset of fault reactivation is observable only when the entire fault is able to slip, i.e., when the slipping patch reaches the sample's edge. Our experimental results suggest that ``drained'' and ``undrained'' conditions lead to two distinct reactivation scenarios. In the ``drained'' case, the slip patch grows behind the fluid pressure front, similarly to previous experimental results obtained at a larger scale \citep{Guglielmi2015}: fault reactivation follows a conventional Coulomb law combined with a simple effective stress law. In the ``undrained'' case, the fault reactivates when only a small fraction of the fault is at elevated fluid pressure. This local increase of fluid pressure induces the propagation of a slip front which largely outgrows the pressurised region. These results are qualitatively consistent with the theoretical analysis developed by \cite{Garagash2012}, whereby a localised fluid pressure increase on a critically loaded fault (here, when $\tau_\mathrm{p}-\tau_0 \ll \mu(P_\mathrm{f,bore} - P_\mathrm{f,0})$, where $P_\mathrm{f,0}$ is the initial homogeneous fluid pressure) acts as a point force. In that case, the pore pressure required to propagate slip is larger than the one expected from Coulomb criterion, and the slipping patch largely outgrows the original pressurised region. Our experiments provide direct evidence for this phenomenon, notably at the highest injection rate where the entire experimental fault is sliding while the fluid pressure front remains localised near the borehole (Figure \ref{fig:wavevel}D). 

In the ``undrained'' scenario, the velocity of the expanding slip patch is expected to be linked to the injection rate \citep[e.g.,][]{Garagash2012,Dublanchet2017}. While the experimental configuration did not allow to capture the propagation of the slip front, our measurements show that the slip rate at the onset of reactivation increases with the injection rate for each stress conditions tested (Figure \ref{fig:overpressure}B). These results suggest that the intensity of the local fluid pressure perturbation enhances the slip-rate and stress transfer at the onset of fault reactivation. Inasmuch as seismogenic fault exhibits generally velocity-weakening behaviour, an increase in the injection rate is expected to enhance the nucleation of instabilities, all the more so than transiently high slip rates can trigger strongly rate-weakening mechanisms that facilitate dynamic rupture propagation. Depending on the fault stress state at large distances from the injection point, slip instabilities can naturally arrest (under low stress conditions) or grow and become large earthquakes (at high stress) \citep[e.g.,][]{Viesca2012,Garagash2012}. 

The hydraulic diffusivity of the fault and surrounding material is expected to increase during slip due to progressive unloading and average decrease of the effective pressure, accelerating the homogenisation of the fluid pressure, as observed in our experiments. This phenomenon is potentially complixified by slip-induced dilatancy, inducing variations in along-fault hydraulic diffusivity. In addition, wear processes, gouge formation and grain crushing is expected to induce a decrease in fault permeability \citep[e.g.,][]{Olsson1993,Zhang1998,Rutter2017}, which might reduce the ability of the fluid pressure front to reach regions far from the injection point (where slip is concentrated, see \cite{Garagash2012}).

Note that in none of our experiments were dynamic stress drops observed during off-loading, although episodes of relatively rapid but stable sliding were seen. This aseismic behaviour is imposed by the stiffness of the apparatus compared to the stiffness of the experimental fault \citep[e.g.,][]{Leeman2016}, and by our experimental procedure, i.e., stress relaxation experiments. In our experimental setup, the dimension of the fault (and its effective compliance) is too small to observe fully developed dynamic runaway ruptures. Faster ruptures might be observable by artificially increasing the compliance of the fault and apparatus system, by imposing a constant applied stress with a servo-controlled load - a rather risky procedure in terms of safety and integrity of the deformation apparatus. However, injection-induced fracture experiments in sandstone indicate that ruptures might be inherently more stable when triggered by fluid pressure increases compared to shear stress increases \citep[e.g.,][]{Ougier2013,Ougier2015,French2016}. 
However, further analysis is required to determine whether such an effect is intrinsic to fluid pressurisation or whether it results from the combination of experimental conditions (e.g., apparatus stiffness) and rock types used.

Taken together, the experimental results presented here emphasize that hydromechanical coupling processes, notably the dependence of transport properties on stress state, have a key control on fault reactivation by localised fluid injection, consistent with theoretical models \citep[e.g.,][]{Cappa2011} and field observations \citep[e.g.,][]{Guglielmi2015}. In nature, measurements of \textit{in situ} stress states in the upper crust have shown that the shear stress is close to the static strength limit for brittle failure \citep[e.g.,][]{Townend2000}. Initial natural stress conditions and fault properties are therefore expected to promote ``undrained'' conditions at the scale of the reservoir, and based on this study, promote the development of fluid pressure heterogeneities along faults during fluid injection. Our experimental results confirm theoretical analyses \citep[e.g.,][]{Garagash2012,Dublanchet2017} showing that these local fluid pressure heterogeneities initiate the propagation of local of slip fronts and induce stress transfer far away from the injection site. The development of such heterogeneities, superimposed with possible preexisting fault structures and background stress profiles, makes accurate predictions of threshold pressure (or injected volume) for fault reactivation and rupture propagation difficult. The time-dependent nature of both fluid flow and fault friction also implies that delayed reactivation is possible (as observed in one of our tests), and changes and spatial variations in fault permeability influence the location and timing of induced seismicity \citep[e.g.,][]{Yeck2016,Chang2016,Vlcek2018}. This implies that in nature, slowing down or stopping fluid injection does not necessarily hamper further fault reactivation, especially as fluids (and fluid pressure) might accumulate along low permeability barriers \citep[e.g.,][]{Yang2017,Passelegue2014}.

\begin{acknowledgments}
FXP acknowledges funding provided by the Swiss National Science Foundation through grant PZENP2/173613. This work was funded, in part, by the UK Natural Environment Research Council through grant NE/K009656/1 to NB. Discussions with Dmitry Garagash, Robert Viesca and Pierre Dublanchet are gratefully acknowledged. Experimental data are available from the UK National Geoscience Data Centre (http://www.bgs.ac.uk/services/ngdc/) or upon request to the corresponding author.
\end{acknowledgments}

\end{article}

\end{document}